\documentclass[review]{elsarticle}
\makeatletter
\def\ps@pprintTitle{%
 \let\@oddhead\@empty
 \let\@evenhead\@empty
 \def\@oddfoot{}%
 \let\@evenfoot\@oddfoot}
\makeatother
\usepackage{lineno,hyperref,graphicx}
\usepackage[normalem]{ulem}
\usepackage[margin=1.0in]{geometry} 
\modulolinenumbers[1]
\setpagewiselinenumbers
\usepackage{multirow}
\usepackage{subfigure}
\usepackage{amsfonts}
\usepackage{amsmath}
\usepackage{booktabs, threeparttable}
\usepackage{array}
\newcolumntype{L}[1]{>{\raggedright\arraybackslash}p{#1}}
\usepackage{color}

\begin{document}

\title{Analysis of ballistic transport and resonance in the $\alpha$-Fermi-Pasta-Ulam-Tsingou model}

\author[ucfaddress] {Nathaniel Bohm}
\author[ucfaddress,AMPACaddress,REACT]{Patrick K. Schelling\corref{mycorrespondingauthor}}
\cortext[mycorrespondingauthor]{Corresponding author.}
\ead{patrick.schelling@ucf.edu}
\address[ucfaddress]{Department of Physics, University of Central Florida, Orlando, FL 32816-2385, USA}
\address[AMPACaddress]{Advanced Materials Processing and Analysis Center, University of Central Florida, Orlando, FL 32816-2385, USA}
\address[REACT] {Renewable Energy and Chemical Transformations (REACT) Cluster, University of Central Florida, Orlando, FL 32816-2385, USA}

\begin{abstract}
Ballistic transport and resonance phenomena are elucidated in the one-dimensional  $\alpha$-Fermi-Pasta-Ulam-Tsingou (FPUT) model using an approach of computing thermal response functions. The existence of periodic oscillations in spatially sinusoidal temperature profiles seen in previous studies is confirmed. However, the results obtained using response functions enable a more complete understanding. In particular, it is shown that resonance involves beats between normal modes which tend to reinforce in a one-dimensional chain. Anharmonic scattering acts to destroy phase coherence across the statistical ensemble, and with increasing anharmonicity, transport is driven towards the diffusive regime. These results provide additional insight into anomalous heat transport in low-dimensional systems. Normal-mode scattering is also explored using time correlation functions. Interestingly, these calculations, in addition to demonstrating loss of phase coherence across an ensemble of simulations, appear to show evidence of so-called q-breathers in conditions of strong anharmonicity. Finally, we describe how the approach outlined here could be developed to include quantum statistics and also also first-principles estimates of phonon scattering rates to elucidate second sound and ballistic transport in realistic materials at low temperatures.
 \end{abstract}


\maketitle

\section{Introduction}
Heat transport in materials is quite often well described by the heat-diffusion equation. However, in one-dimensional systems, heat conduction is often found to be ``anomalous''. In these cases, the apparent thermal conductivity is found to depend on system size $L$ and may diverge as $\kappa(L) \sim L^{\gamma}$ with $0<\gamma < 1$\cite{Benenti_2020,Lepri:1997aa,Lepri_1998}.  In a somewhat broader context, divergences of this kind were first suggested  by Alder and Wainright\cite{Wainwright:1971aa,Alder:1967aa}.  Using Kubo expressions to compute thermal conductivity, divergences are observed as power-law decays in current-current correlation functions, $\langle J(t) J(0) \rangle \sim t^{\gamma-1}$ \cite{Benenti_2020,POMPEAU:1975aa}. In one-dimensional chains based on the Fermi-Pasta-Ulam-Tsingou (FPUT) model, divergent behavior has been reported with $\gamma={1 \over 3}$\cite{Dhar_2008}. Somewhat more recent results for FPUT lattices using thermostats and explicit temperature gradients has suggested a mixture of ballistic and diffusive transport and divergent thermal conductivity with $\gamma={2 \over 5}$\cite{Dematteis:2020fm}. Experimental evidence has also demonstrated the breakdown of Fourier's law in carbon and boron-nitride nanotubes\cite{Chang:2008wd}.
 Experimental and computational results for single-layer suspended graphene \cite{Xu_2014} indicate divergence in $\kappa \sim \log L$. Theoretical calculations of second sound in suspended graphene have also been reported\cite{Lee_2015,Li:2019vo}. Even more recently, second sound has been reported in experimental studies of graphite\cite{Ding_2022}. However, while the divergence of $\kappa(L)$ with system size is generally understood to be due to a breakdown in the assumption of diffusive transport, investigation of anomalous thermal transport still remains an active area. 
 
 The current understanding of anomalous thermal conduction clearly implicates the role of ballistic transport. It might then be proposed that inquiry based on understanding size-dependent thermal conductivity $\kappa(L)$ is rather limiting. Specifically, while measurement of $\kappa(L)$ in an experiment is of obvious practical and fundamental importance, the very fact that $\kappa(L)$ depends on system size indicates that the heat diffusion equation itself does not apply. One might reasonably question how obtaining the scaling behavior of $\kappa(L)$ could lead to more fundamental understanding of anomalous thermal transport without developing better understanding of ballistic transport.
 
 In this paper, we take the perspective that computational and theoretical approaches should be developed to elucidate transport without making a priori assumptions as to whether the heat diffusion equation provides a suitable description. In the context of Kubo theory, this suggests that current-current correlation functions should be used to elucidate transport without the assumption that Fourier's law applies. Instead, Kubo theory might be used to compute transport in a ballistic regime or even a regime intermediate between purely ballistic and diffusive.  In addition, we make note that previous studies focus on steady-state investigations to determine $\kappa(L)$. By contrast, experiments and applications are often interested in time-dependent phenomena, where transport might involve not only heat diffusion but also heat waves\cite{Joseph:1989ud} and second sound\cite{Chester:1963vm}. For example, quite recent work has demonstrated the existence of second sound up to room temperature in bulk germanium driven by rapidly-varying heat inputs from a pump-probe laser system\cite{Beardo_2021}. In this experimental work, heat transport beyond the diffusive regime was demonstrated by measurement of time-dependent phenomena which could obviously not be characterized by the heat diffusion equation.

With these considerations in mind, this paper presents application of recently developed computational methodology for computation of thermal response functions\cite{Fernando_2020} to elucidate transport phenomena in one-dimensional FPUT chains. A similar approach based on computing the response to externally-applied perturbations was reported recently for one-dimensional chains\cite{Allen:2022we}. The advantage of the approach is that the response functions do not require any assumptions about whether transport is ballistic or diffusive. Moreover, the response functions describe the time-dependent response to heat pulses, and therefore are relevant for elucidating second sound. 
In the regime where diffusive transport is relevant, the thermal response functions begin to approach behavior closely consistent with the heat diffusion equation\cite{Fernando_2020}. In this paper, we apply this approach to understanding heat transport in the $\alpha$-FPUT model. Recently, time-dependent ballistic resonance has been observed in $\alpha$-FPUT models\cite{Kuzkin:2020tr}. In ballistic resonance, oscillations are anharmonically coupled to ``mechanical'' vibrations which can grow in time.  By contrast, $\beta$-FPUT chains exhibit resonance and oscillatory transport behavior without coupling to mechanical vibrations, and hence do not display ballistic resonance\cite{Korznikova:2020te}.
The oscillatory behavior has been modeled using continuum differential equations \cite{Kuzkin:2020tr,Krivtsov_2015}. In this paper, we use response functions to provide further insight into ballistic transport and resonant behavior without directly addressing the coupling to mechanical vibrations inherent to ballistic resonance. We demonstrate the importance of phase coherence between harmonic normal modes, with the resonant behavior due to reinforcement between beat frequencies across the entire normal-mode spectrum. This picture provides insight into oscillatory behavior in FPUT chains reported earlier\cite{Kuzkin:2020tr,Korznikova:2020te}. We show here that resonance is primarily due to harmonic wave interference, with anharmonicity acting primarily to destroy phase coherence. The results obtained demonstrate that reinforcing beat frequencies is an important factor in observing ballistic transport that is unique to one-dimensional systems. Hence, the results here may provide additional theoretical insight into anomalous thermal transport and diverging thermal conductivity in low-dimensional systems.

\section{Theory and Methodology}

We consider a linear chain of $N$ particles, with each particle coupled to its nearest neighbors. Periodic-boundary conditions are applied.
The potential energy corresponds to the $\alpha$-FPUT model. Specifically the potential energy function $V(x)$ is given by,
\begin{equation}
V(x)={1 \over 2} x^{2} + {1 \over 3} \alpha x^{3}
\end{equation}
in which $x=r_{n}-r_{n-1}$ and $r_{n}$ represents the displacement from equilibrium.
The equation of motion for each particle is, taking $m=1$ for the mass,
\begin{equation}
\ddot{r}_{n} = F_{n+1}+F_{n-1}
\end{equation}
where $F_{n \pm 1}= - V^{\prime} (r_{n}-r_{n \pm 1})$.
The local heat current at site $n$ is given by,
\begin{equation} \label{current}
J_{n}= {1 \over 2}  \left( F_{n-1}-F_{n+1}\right) p_{n}
\end{equation}
in which $p_{n}=m \dot{r}_{n}$ with $m=1$.

Following our previous work\cite{Fernando_2020}, the response function $K_{mn}$ describes the heat current response due to an external heat source $u^{(ext)}(t)$,
\begin{equation} \label{nl}
J_{n}(t) = -{1 \over 2N} \sum_{m=1,N} K_{nm}(t-t^{\prime}) \left[ u^{(ext)}_{m+1}(t^{\prime}) - u^{(ext)}_{m-1}(t^{\prime}) \right] .
\end{equation}
This expression will also be written in reciprocal space.  The heat current and external source are written in terms of a discrete Fourier series. Assuming an even number of sites $N$, this leads to the expressions for the local heat current,
\begin{equation} 
J_{n}(t)= \sum_{q=-{N \over 2}}^{{N \over 2}-1} \tilde{J}_{q}(t)  e^{2 \pi i n q \over N}
\end{equation}
and the source term,
\begin{equation} 
u^{(ext)}_{m}(t^{\prime})= \sum_{q=-{N \over 2}}^{{N \over 2}-1} \tilde{u}^{(ext)}_{q}(t^{\prime})  e^{2 \pi i mq \over N} . 
\label{ext}
\end{equation}
In these equations and hereafter, tildes are used to identify reciprocal space quantities.

The response function itself can be written as a Fourier series due to the periodic nature of the chain. Hence we have
\begin{equation} 
 K_{nm}(t-t^{\prime})  = \sum_{q=-{N \over 2}}^{{N \over 2}-1} \tilde{K}_{q}(t-t^{\prime})e^{2 \pi i (n-m)q \over N} .
\end{equation}
Then the response equation is written in reciprocal space,
\begin{equation} 
 \tilde{J}_{q}(t)=-i \sin\left({2 \pi q \over N}\right)\tilde{K}_{q}(t-t^{\prime})\tilde{u}^{(ext)}_{q}(t^{\prime}) .
 \label{linresp}
\end{equation}

Here we provide a brief review of the derivation of response function approach, although more details can be found in our previous work\cite{Fernando_2020}. Following the assumptions made by Onsager and Kubo, it should be the case that equilibrium fluctuations can be related to the dissipation of an external perturbation. Hence, from Eq. \ref{linresp}, we associate the external source with a fluctuation in the equilibrium ensemble. Taking $t^{\prime}=0$, we multiply Eq. \ref{linresp} by $i\tilde{u}_{-q}(0)$ and take an ensemble average,
\begin{equation} 
i \langle \tilde{J}_{q}(t)\tilde{u}_{-q}(0) \rangle = \sin\left({2 \pi q \over N}\right)\tilde{K}_{q}(t) \langle \tilde{u}_{q}(0)\tilde{u}_{-q}(0) \rangle .
\end{equation}
The final equation is obtained by using time-reversal symmetry and differentiating with respect to time. In addition, the continuity equation in reciprocal space is used,
\begin{equation}
{\partial \tilde{u}_{q} \over \partial t} = -i \sin{\left({2 \pi q \over N}\right)} \tilde{J}_{q} .
\end{equation}
The final expression then relates the time-correlation functions to the response function,
\begin{equation}\label{compresp}
\tilde{K}_{q}(\tau)= {\int_{0}^{\tau} \langle \tilde{J}_{q}(t) \tilde{J}_{-q}(0)  \rangle dt  \over  \langle  \tilde{u}_{q}(0) \tilde{u}_{-q}(0)\rangle } .
\end{equation}
The significance of this equation is that fluctuations in the current computed within the equilibrium ensemble can be used to determine the response function $\tilde{K}_{q}(\tau)$, which subsequently can be used to compute the response of a system to heat inputs due to external sources as long as the system is maintained in the linear-response regime. In this sense the approach is comparable to the usual application of Green-Kubo theory to compute thermal conductivity. However, the approach above makes no assumptions about the applicability of Fourier's law, and hence it is useful for describing both diffusive and ballistic transport.

Additional insight into the FPUT problem can be obtained by working in a basis of
normal-mode coordinates $Q_{k}$ and $P_{k}$ defined by,
\begin{equation}
Q_{k}={1 \over \sqrt{N}} \sum_{n=0}^{N-1} r_{n} e^{-{2 \pi i nk \over N}}
\end{equation}
\begin{equation}
P_{k}={1 \over \sqrt{N}} \sum_{n=0}^{N-1} p_{n} e^{-{2 \pi i nk \over N}}
\end{equation}
with $p_{n}= \dot{r}_{n}$. Using this representation, the total energy $E_{tot}$ is given as a summation over the contributions due to each of the $N-1$ normal modes,
\begin{equation}
\label{etot}
E_{tot} =  \sum_{k=-{N \over 2}+1}^{N \over 2}  E_{k}=\sum_{k=-{N \over 2}+1}^{N \over 2}  \left( {P_{k}P_{-k} \over 2} + {1 \over 2} \omega_{k}^{2} Q_{k}Q_{-k} \right)
\end{equation}
in which the energy of the anharmonic term in the potential energy is not included. The dispersion relation for the normal modes is given by,
\begin{equation} 
\omega_{k}^{2} =4 \sin^{2} \left ({\pi k \over N} \right )  .
\end{equation}

It has been shown\cite{Onorato:2015aa} that the rate of thermalization is determined by the parameter,
\begin{equation} \label{init}
\epsilon=\alpha \sqrt{{E_{tot}\over N-1}}
\end{equation}
in which $E_{tot}$ is the total energy given by Eq. \ref{etot}. In the cases of very small chains with $N=16$ and $N=32$,  the timescale for equilibration scales $1/\epsilon^{8}$ \cite{Onorato:2015aa}. In the following, the initial states are always taken to satisfy classical equipartition but with random initial phases for each normal mode.  To explore the effect of anharmonicity on the response functions, we report results for different values of $\epsilon$. The parameter $\alpha=0.15$ is always used. Hence, different values of $\epsilon$ correspond to different initial amplitudes according to Eq. \ref{init}. While the initial states should always correspond to thermal equilibrium, an initial run period of $10^{4}$ steps was computed before any statistical averaging was implemented.

In addition to heat currents computed exactly according to Eq. \ref{current}, the approximate heat current obtained only from the harmonic forces is also computed.
It can be shown that in this approximation the heat current is given in terms of the normal-mode coordinates $Q_{k}$ and $P_{k}$ as,
\begin{equation}
J_{n} =  -{1 \over N} i \sum_{k} \sum_{l} \sin{\left( {2 \pi k \over N}\right)} Q_{k}P_{-l}e^{2 \pi i n(k-l) \over N} .
\end{equation}
The Fourier components $\tilde{J}_{q}$ of this expression are also determined. Specifically, we note that $\tilde{J}_{+q}$ is defined by,
\begin{equation}
\tilde{J}_{+q}= -{1 \over N^{2}}i \sum_{n} \sum_{k} \sum_{l}  \sin{\left( {2 \pi k \over N}\right)}   Q_{k}P_{-l}e^{2 \pi i n(k-l-q) \over N} .
\end{equation}
The summation on sites $n$ yields zero unless $k-l-q=0$, or alternately $l=k-q$. Then after evaluating the sums on $n$ and $l$,
\begin{equation}\label{currpq}
\tilde{J}_{+q}= -{1 \over N}i\sum_{k } \sin{\left( {2 \pi k \over N}\right)}   Q_{k}P_{-k+q} .
\end{equation}
Similarly we can write,
\begin{equation}\label{currmq}
\tilde{J}_{-q}= {1 \over N}i\sum_{k }\sin{\left( {2 \pi k \over N}\right)}   Q_{-k}P_{k-q} .
\end{equation}
These expressions will also be used to compute response functions. As will be shown, this representation is extremely revealing for establishing the role of different normal modes in the response. It is also helpful in demonstrating that mode coherence and interference effects are responsible for resonance and ballistic transport.

We next obtain an expression for $\tilde{u}_{q}$ in the harmonic approximation in terms of the normal-mode coordinates. Assuming potential energy is equally shared between connected particles, the local energy $u_{n}$ is defined by,
\begin{equation}
u_{n}={1 \over 2} p_{n}^{2} + {1 \over 4} \left(r_{n}-r_{n-1} \right)^{2} +{1 \over 4} \left(r_{n+1}-r_{n} \right)^{2} 
\end{equation}
in which only the harmonic interactions are included.
Then the Fourier component $\tilde{u}_{q}$ is given by,
\begin{equation}
\tilde{u}_{q} = {1 \over N} \sum_{n} u_{n} e^{-{2 \pi iqn \over N}} .
\end{equation}
It is then straightforward to show that,
\begin{equation}
\tilde{u}_{q} = {1 \over 2N }\sum_{k} P_{k} P_{-k+q}
+{2 \over N}\sum_{k} Q_{k} Q_{-k+q} \cos{\left({\pi q \over N}\right)} \sin{\left({\pi k \over N}\right)} \sin{\left({\pi (k-q) \over N}\right)} .
\label{uharm}
\end{equation}
For the case $q=0$, we obtain 
\begin{equation}
\tilde{u}_{q=0} = {1 \over N }\sum_{k}\left[ {P_{k} P_{-k} \over 2} 
+{1 \over 2}\omega_{k}^{2} Q_{k} Q_{-k} \right] .
\end{equation}
Comparison with Eq. \ref{etot} shows that $\tilde{u}_{q=0}={E_{tot} \over N}$ is the average total energy per particle. 


Following other authors, we used the SABA$_2$C symplectic integrator first introduced in Ref. \cite{Laskar_2001} with a time step $dt=0.10$ \cite{Pace_2019}. Tests of the integrator for $N=32$ chains demonstrated thermalization times in very close agreement to the results from Ref. \cite{Onorato:2015aa}, including validation of the ${1 \over \epsilon^{8}}$ dependence of the thermalization time. The simulations reported here use longer chains with $N=512$ and $N=16,384$ sites. 

Initial conditions corresponded to an equal amount of energy in each normal mode. Specifically then, the initial condition for each ensemble member satisfies the expression,
\begin{equation}
 |P_{k}|^{2} =  \omega_{k}^{2} |Q_{k}|^{2} = {E_{tot} \over N-1}
\end{equation}
for each normal mode $k$. 
To randomize the initial conditions, the phases of each normal mode were selected randomly. Then the initial conditions for the displacements $r_{n}$ and momenta $p_{n}$ are given by summations over normal modes $k$,
\begin{equation}
r_{n} ={1 \over \sqrt{N}} \sum_{k}|Q_{k}| \cos\left({2 \pi nk \over N} - \phi_{k}\right)
\end{equation}
\begin{equation}
p_{n} ={1 \over \sqrt{N}}  \sum_{k}|Q_{k}| \omega_{k} \sin\left({2 \pi nk \over N} - \phi_{k}\right) = {1 \over \sqrt{N}} \sum_{k}|P_{k}| \sin\left({2 \pi nk \over N} - \phi_{k}\right) 
\end{equation}
in which $\phi_{k}$ is the random phase.
 In each instance, reported quantities correspond to ensemble averages with the number of independent ensemble members indicated. In the case of response functions, time averaging over each member of the ensemble is also used.  Finally, results for $\epsilon$ above $0.08$ are not reported. It was found that due to the cubic anharmonicity, the FPUT lattice was unstable to occasional collapse for $\epsilon>0.08$. Consequently, it was not possible to observe transition to diffusive transport. By contrast, it has been  previously demonstrated that simulations of the $\beta$-FPUT model\cite{Korznikova:2020te} can be extended to stronger anharmonic regimes where ballistic transport and resonance is not evident. Here, we were not able to explore this transition.

\section{Results}
\begin{figure}
\begin{centering}
\includegraphics[width=0.5\textwidth]{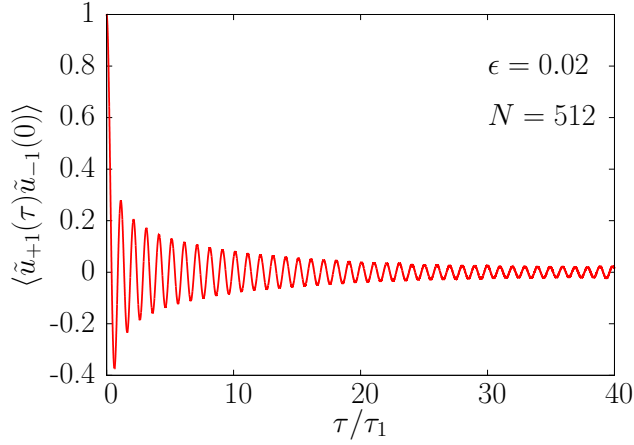} 
\caption{Resonant behavior shown by the equilibrium correlation function $\langle \tilde{u}_{+1}(\tau) \tilde{u}_{-1}(0) \rangle$ for simulations with $\epsilon=0.02$ and $N=512$. Ensemble averaging was performed for $1280$ independent runs each with $1.5 \times 10^{6}$ MD steps. Results normalized by the average fluctuation $\langle \tilde{u}_{+1}(0) \tilde{u}_{-1}(0) \rangle$.
}
\label{uu0.02}
\end{centering}
\end{figure}

\begin{figure}
\begin{centering}
\includegraphics[width=0.5\textwidth]{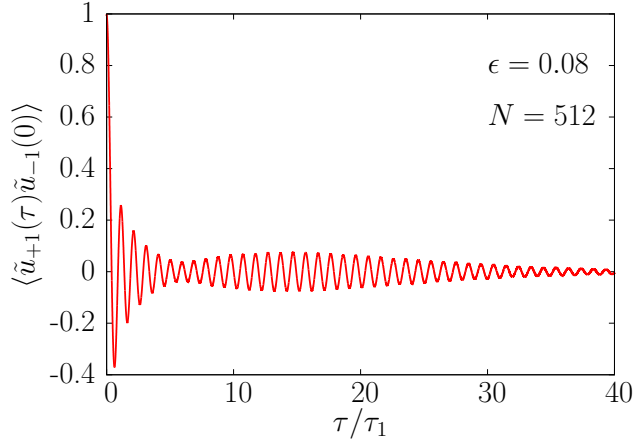} 
\caption{Resonant behavior shown by the equilibrium correlation function $\langle \tilde{u}_{+1}(\tau) \tilde{u}_{-1}(0) \rangle$ for simulations with $\epsilon=0.08$ and $N=512$. Ensemble averaging was performed for $1280$ independent runs each with $1.5 \times 10^{6}$ MD steps. Results normalized by the average fluctuation $\langle \tilde{u}_{+1}(0) \tilde{u}_{-1}(0) \rangle$.
}
\label{uu0.08}
\end{centering}
\end{figure}

For all of the simulations conditions reported here, oscillatory behavior was found in agreement with previous studies \cite{Kuzkin:2020tr,Korznikova:2020te}. Resonant behavior is most directly shown by plotting the ensemble-averaged energy fluctuations
$\langle \tilde{u}_{1}(\tau)\tilde{u}_{-1}(0)\rangle$. The quantities $\tilde{u}_{q}$ are defined in terms of the normal mode coordinates by Eq. \ref{uharm}. In Fig. \ref{uu0.02}, the energy fluctuations are plotted for a chain with $N=512$ particles and $\epsilon=0.02$. The period of the oscillations corresponds closely to the period $\tau_{1} = {\pi \over \sin{\left({\pi \over N}\right)}}$ for the normal mode $k=1$. Harmonic behavior persists to long times even for $\epsilon=0.08$ as shown in Fig. \ref{uu0.08}. Increased anharmonicity for $\epsilon=0.08$ leads to more rapid decay of the oscillations. This point will be displayed more convincingly in relation to the response functions.

\begin{figure}
\begin{centering}
\includegraphics[width=0.5\textwidth]{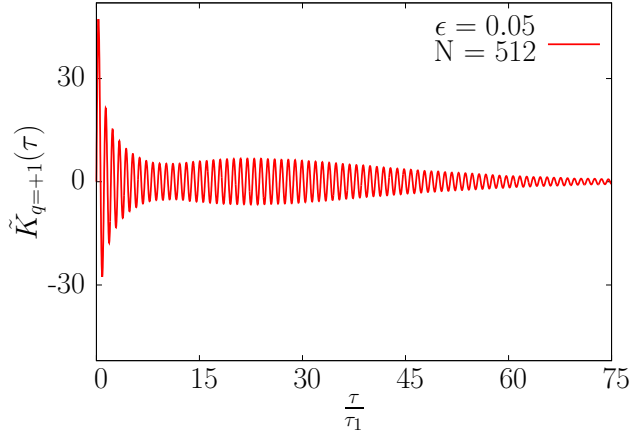} 
\caption{ 
The response function $\tilde{K}_{q=+1}(\tau)$ obtained for $\epsilon=0.05$ and a system with $N=512$ particles. The response function was computed from an ensemble of 1280 independent runs, with each ensemble member comprised of $5 \times 10^{6}$ MD steps. The time $\tau$ is scaled by the longest vibrational period in the system $\tau_{1}={\pi \over \sin{\pi \over N}}$. 
}
\label{K0.05}
\end{centering}
\end{figure}

\begin{figure}
\begin{centering}
\includegraphics[width=0.5\textwidth]{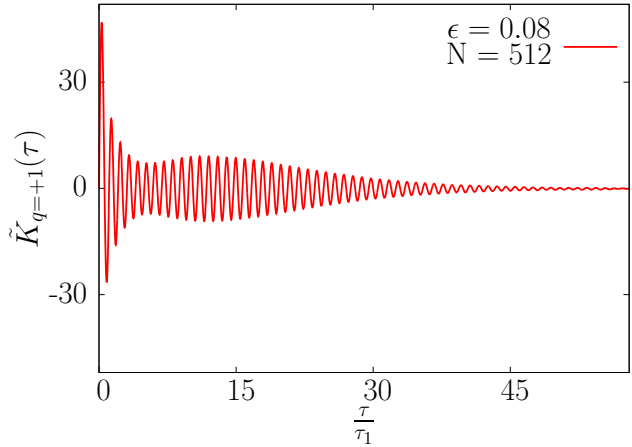} 
\caption{
The response function $\tilde{K}_{q=+1}(\tau)$ obtained for $\epsilon=0.08$ and a system with $N=512$ particles. The response function was computed from an ensemble of 1280 independent runs, with each ensemble member comprised of $5 \times 10^{6}$ MD steps. The time $\tau$ is scaled by the longest vibrational period in the system $\tau_{1}={\pi \over \sin{\pi \over N}}$. 
}
\label{K0.08}
\end{centering}
\end{figure}

Ballistic resonant behavior is also exhibited in the response function $\tilde{K}_{q=+1}(\tau)$. In Fig. \ref{K0.05} and Fig. \ref{K0.08}, the computed response functions are shown for $N=512$ chains using the exact heat current including anharmonic terms. As with the energy fluctuations, resonant behavior with period $\tau \approx \tau_{1}$ is demonstrated. 
Comparison of results for $\tilde{K}_{q=+1} (\tau)$ for different values of $\epsilon$ demonstrates the primary role of anharmonicity. Specifically, for smaller values of $\epsilon$, phase coherence and wavelike response persists for longer times. As $\epsilon$ is increased, there is an increased decay rate for the oscillatory behavior.  For $\epsilon=0.08$ shown in Fig. \ref{K0.08}, the oscillations have decayed essentially completely by $\sim 50$ periods. By contrast, for  $\epsilon=0.05$ shown in Fig. \ref{K0.05}, oscillatory behavior is still perceptible after $\sim 75$ periods. This general trend was reported elsewhere for simulations of the $\beta$-FPUT model\cite{Korznikova:2020te}. However, in the results reported here, diffusive behavior is never observed.

\begin{figure}
\begin{centering}
\includegraphics[width=0.5\textwidth]{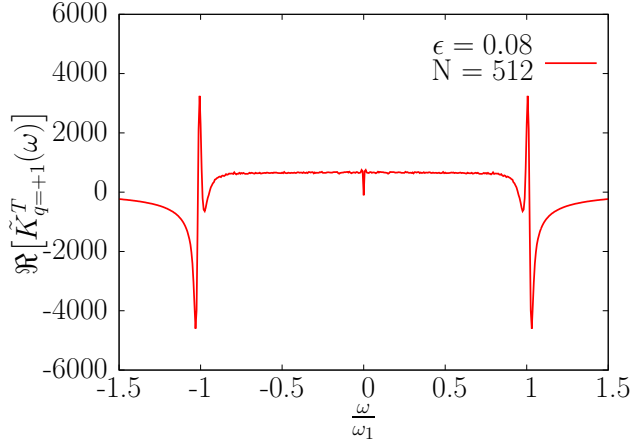} 
\caption{
Real part of the Fourier transform $\Re{\left[\tilde{K}_{q=+1}^{T}(\omega)\right]}$ obtained for $\epsilon=0.08$ and $N=512$. Predominant oscillatory behavior is shown as peaks at frequencies $\omega \approx \omega_{1}$. 
}
\label{RFT}
\end{centering}
\end{figure}

\begin{figure}
\begin{centering}
\includegraphics[width=0.5\textwidth]{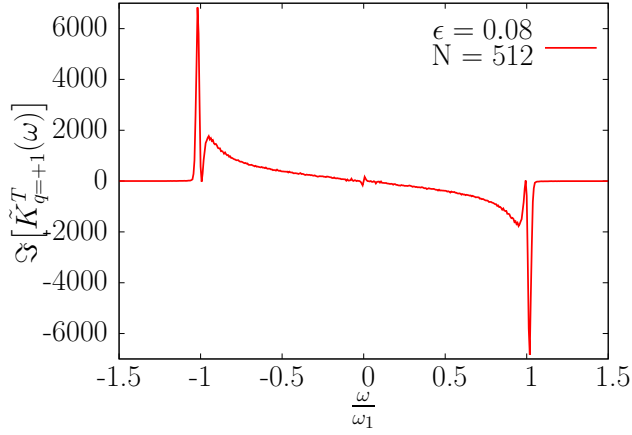} 
\caption{
Imaginary part of the Fourier transform  $\Im{\left[\tilde{K}_{q=+1}^{T}(\omega)\right]}$ for $\epsilon=0.08$ and $N=512$. Predominant oscillatory behavior is shown as peaks at frequencies $\omega \approx \omega_{1}$. 
}
\label{IFT}
\end{centering}
\end{figure}

Fourier transform of the response function $\tilde{K}_{q=+1} (\tau)$ using the equation,
\begin{equation}
\tilde{K}_{q=+1}^{T}(\omega) = \int_{0}^{T_{max}} \tilde{K}_{q=+1} (\tau) e^{-i \omega \tau} d \tau
\end{equation}
demonstrates that the primary frequency in the response corresponds to the lowest normal-mode frequency $\omega_{1}=2 \sin{\pi \over N}$. However, additional frequencies both faster and slower that $\omega_{1}$ are apparent. As shown in the Analysis section, these details can be linked specifically to the normal-mode dispersion $\omega_{k}^{2}=4 \sin^{2}\left({\pi k \over N}\right)$ and to the finite lifetime for coherent oscillation due to anharmonicity. Hence, although the response is pronounced at frequency $\omega_{1}$, in fact the entire normal-mode spectrum plays an important role in the response function. Typical results are shown in Fig. \ref{RFT} and Fig. \ref{IFT} for the real and imaginary parts of $\tilde{K}_{q=+1}^{T}(\omega) $ for simulations with $\epsilon=0.08$ and $N=512$. Very sharp peaks are evident for $\omega \approx \omega_{1}$ and $\omega \approx -\omega_{1}$. These results further demonstrate wave-like transport and resonant behavior even for the largest value of $\epsilon$ simulated here.

\begin{figure}
\begin{centering}
\includegraphics[width=0.5\textwidth]{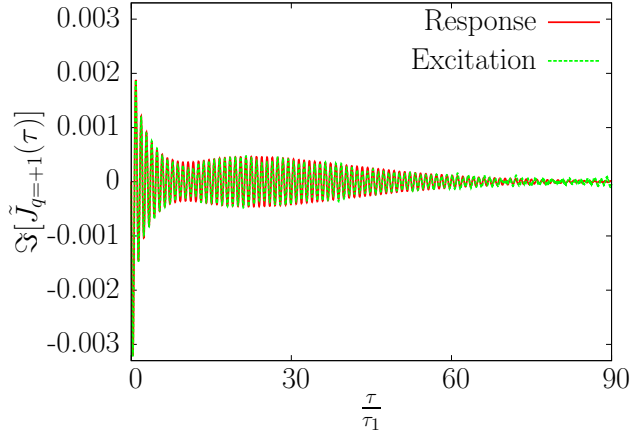}
\caption{
The imaginary part of the heat current $\Im \left[ \tilde{J}_{q=+1}(\tau)\right]$ resulting from direct excitation based on Eq. \ref{excite} with $\delta=0.20$ and $\epsilon=0.05$ for a system with $N=512$ particles. Comparison is made to the prediction based on the computed response function $\tilde{K}_{q=+1}(\tau)$. The response function was computed from an ensemble of 3072 independent runs. Response functions were also time-averaged over $5 \times 10^{6}$ MD steps. Direct excitation results were averaged over 6144 simulations with independent initial conditions.
}
 \label{validate}
\end{centering}
\end{figure}

\begin{figure}
\begin{centering}
\includegraphics[width=0.5\textwidth]{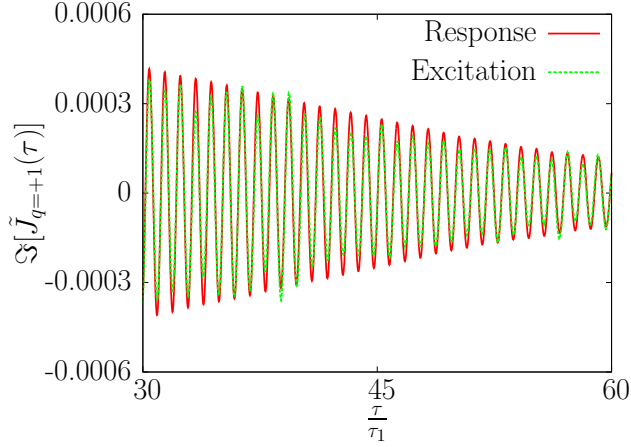}
\caption{
The imaginary part of the heat current $\Im \left[ \tilde{J}_{q=+1}(\tau)\right]$ resulting from direct excitation based on Eq. \ref{excite} with $\delta=0.20$ and $\epsilon=0.05$ for a system with $N=512$ particles. Comparison is made to the prediction based on the computed response function $\tilde{K}_{q=+1}(\tau)$. 
}
 \label{validate2}
\end{centering}
\end{figure}

The validity of the response functions will next be demonstrated by comparison with calculated heat currents that result from an explicit perturbation of an equilibrium system. The demonstration corresponds to a perturbation with $q=\pm 1$ in Eq. \ref{ext}. In detail, the velocities $\dot{r}_{n}(t=0)$ are scaled according to,
\begin{equation} \label{excite}
\dot{r}_{n}(0) \rightarrow \dot{r}_{n}(0) \sqrt{1 +\delta \cos{\left({2 \pi n \over N}\right)}}. 
\end{equation}
Also we recall here the relationship $p_{n}=\dot{r}_{n}$.
Figure \ref{validate} compares the current resulting from the direct excitation with the prediction obtained using the response function  $\tilde{K}_{q=+1}(\tau)$ obtained for $\epsilon=0.05$. Details for the total number of independent ensemble members and MD integration steps are included in the captions. The amplitude of the excitation is given by $\delta=0.20$. The predicted resonant behavior is in very good agreement, indicating the validity of the response function. The oscillations persist for very long times. In Fig. \ref{validate2}, the same calculations are compared only for times $30 \leq {\tau \over \tau_{1}} \leq 60$ to show more clearly the close agreement at even longer times. Statistical error was found to be most sensitive to the number of direct excitation calculations, since these are only averaged over independent initial conditions and are not time-averaged like response functions.  Agreement between direct excitation and the predictions based on the response functions improved by increasing the number of direct excitation calculations to 6144 independent simulations, in contrast to an ensemble of 3072 independent calculations to determine the response function.

\begin{figure}
\begin{centering}
\includegraphics[width=0.5\textwidth]{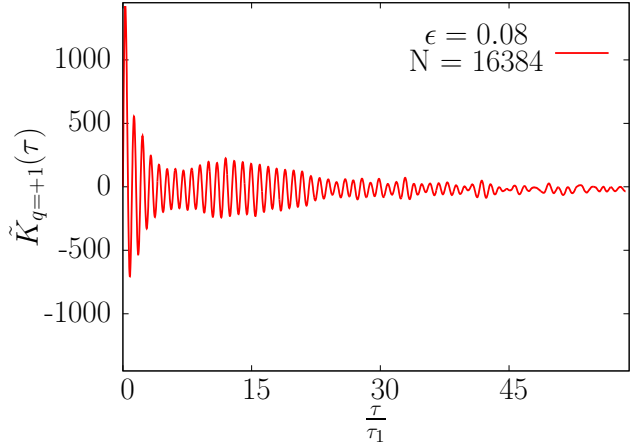} 
\caption{
The response function $\tilde{K}_{q=+1}(\tau)$ obtained for $\epsilon=0.08$ and a system with $N=16384$ particles. The response function was computed from an ensemble of 512 independent runs, with each ensemble member comprised of $1 \times 10^{7}$ MD steps. The time $\tau$ is scaled by the longest vibrational period in the system $\tau_{1}={\pi \over \sin{\pi \over N}}$. 
}
\label{bigK}
\end{centering}
\end{figure}

Having established the resonant behavior for $N=512$ particle chains along with a demonstration of the validity of the response functions, we next present results for much longer chains with $N=16,384$ particles. The only clear difference for $\tilde{K}_{q=+1}(\tau)$ between $N=512$ and $N=16,834$ chains is in the primary resonant frequency. In both cases, the resonant behavior corresponds to oscillations with frequency $\omega \approx \omega_{1}$, and period $\tau \approx \tau_{1}$. In Fig. \ref{bigK}, $\tilde{K}_{q=+1}(\tau)$ is again plotted as a function of $\tau \over \tau_{1}$ for $N=16384$ and $\epsilon =0.08$. Comparison to the same conditions for the $N=512$ chain in Fig. \ref{K0.08} shows no obvious differences. The only apparent feature which is distinct is the presence of more statistical noise for the large chain due to the relatively smaller ensemble for the large chain. In particular, while the total simulation time in Fig. \ref{bigK} was increased to $10^{7}$ steps, the oscillation period is longer by a factor of 32. Moreover, due to the increased computational cost, only $512$ independent initial conditions were used to generate the data plotted in Fig. \ref{bigK}. Nevertheless, it is apparent that as long as times are scaled by the increased period associated with the longer wavelength in the $N=16,384$ chain, at least in this range the dominant behavior is apparently independent of chain length. However, there are reasons to posit that this situation may not persist as chain length is increased which will be discussed in the next section.

In summary, the predominant feature exhibited by the response function  $\tilde{K}_{q=+1}(\tau)$ is resonant behavior at frequency $\omega \approx \omega_{1}$, where $\omega_{1} = 2\sin {\left({\pi \over N}\right)}$ is the frequency associated with the longest wavelength mode allowed in the periodic chain. Increasing chain length $N$ appears to only have the effect of  increasing the observed period for resonant behavior.  In the next section, it will be demonstrated that resonant behavior is an entirely harmonic phenomenon caused by beats between different normal modes. 
The amplitude of the oscillations in $\tilde{K}_{q}(\tau)$ tends to decay with time $\tau$. There are two origins for this behavior. First, as the next section will demonstrate, oscillations depend on phase coherence and interference between normal modes. However, because mode dispersion is nonlinear in the FPUT model, the beat frequencies differ across the spectrum. Over time, phase differences accumulate between the beats resulting in a decrease in the amplitude of $\tilde{K}_{q}(\tau)$. The second reason for the decrease in the oscillatory behavior exhibited by $\tilde{K}_{q}(\tau)$ is phase decoherence caused by anharmonicity. This effect can be elucidated by simulations using different values of the parameter $\epsilon$. It is of interest specifically to establish how phase decoherence depends on $\epsilon$, and moreover whether oscillations can be damped for large enough values of $\epsilon$ such that diffusive transport becomes relevant.

\section{Analysis}

Analysis of the results is based on correlations determined from only the harmonic form of the heat current given in Eqs. \ref{currpq}-\ref{currmq}. Using the harmonic currents, the current-current correlation is given by the expression,
\begin{equation} \label{corrharm}
\langle \tilde{J}_{+q}(\tau) \tilde{J}_{-q}(0)\rangle={1 \over N^{2}}\sum_{k }\sum_{k^{\prime} } \sin{\left( {2 \pi k \over N}\right)}  \sin{\left( {2 \pi k^{\prime} \over N}\right)} 
\langle Q_{k}(\tau)P_{-k+q}(\tau)  Q_{-k^{\prime}}(0)P_{k^{\prime}-q}(0)\rangle .
\end{equation}
Written in this form the origin of the beat phenomenon is very clear. Specifically, the dominant terms in Eq. \ref{corrharm} correspond to those with $k=k^{\prime}$ and hence involve modes $k$, $-k$, $-k+q$, $k-q$. As a result, the most important frequencies involve the difference $\omega_{k}-\omega_{k \pm q}$. When $k$ and $k-q$ are small enough to consider the normal-mode dispersion to be linear, $\omega_{k}-\omega_{k \pm q} \approx \omega_{q}$. Hence, as long as $q$ corresponds to a long-wavelength fluctuation, many modes contribute to resonant behavior with frequency $\omega_{q}$. While it may seem natural to assume that time dependence with frequency $\omega_{q}$ involves primarily the normal mode $q$, this is shown clearly not to be the case. In fact, as shown below, many modes contribute to resonant transport.

To analyze the interference effects and the coherent beat phenomenon, we examined contributions to Eq. \ref{corrharm} due to specific sets of normal-mode coordinates which exhibit a single beat frequency. We specifically consider contributions to Eq. \ref{corrharm},
\begin{equation} 
F_{q,k}(\tau)={1 \over N^{2}} \sin{\left( {2 \pi k \over N}\right)}    \langle \left[Q_{k}(\tau)P_{-k+q}(\tau) + Q_{-k+q}P_{k}(\tau) \right] \sum_{k^{\prime}} \sin{\left( {2 \pi k^{\prime} \over N}\right)}Q_{-k^{\prime}}(0)P_{k^{\prime}-q}(0)\rangle .
\end{equation}
We define the partial response function,
\begin{equation} 
\tilde{K}^{P}_{q,k}(\tau)={ \int_{0}^{\tau} F_{q,k}(t)dt \over \langle \tilde{u}_{q}(0) \tilde{u}_{-q}(0) \rangle}
\end{equation}
which is related to the complete response function in Eq. \ref{compresp} by $\tilde{K}_{q}(\tau) =  \sum_{k} \tilde{K}^{P}_{q,k}(\tau)$.
Calculation of $\tilde{K}^{P}_{q,k}(\tau)$ serves to isolate contributions due to discrete parts of the spectrum.

\begin{figure}
\begin{centering}
\includegraphics[width=0.5\textwidth]{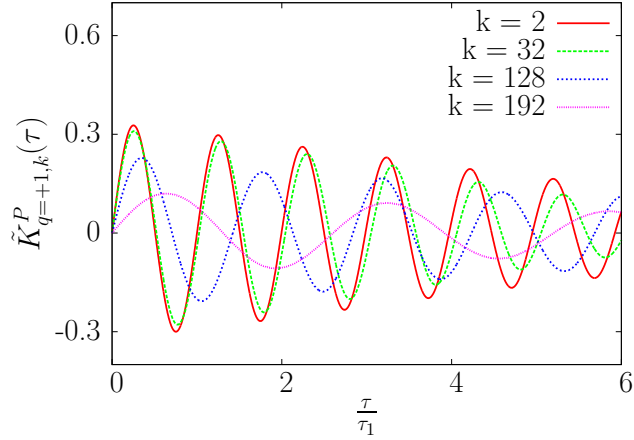} 
\caption{
Partial response function $\tilde{K}^{P}_{q=+1,k}(\tau)$ obtained for $\epsilon=0.08$ and a system with $N=512$ particles. Results are presented for $k=2$, $k=32$, $k=128$, and $k=192$ as shown in the legend. The time $\tau$ is scaled by the longest vibrational period in the system $\tau_{1}={\pi \over \sin{\pi \over N}}$. These results were obtained from an ensemble of $768$ independent simulations with $1.5 \times 10^{6}$ MD steps in each. The same ensemble details apply to Figs. 11-14.
}
\label{K*1}
\end{centering}
\end{figure}

We provide an analysis of $\tilde{K}^{P}_{q,k}(\tau)$ for the $N=512$ particle chain. In Fig. \ref{K*1}, results are shown for a few different modes $k$ for very short correlations times. Several interesting features emerge from this analysis. Contributions due to long wavelength modes exhibit a beat frequency comparable to $\omega_{1}$. Consequently, many modes contribute coherent resonant behavior that reinforces constructively. In Fig. \ref{K*1}, this is apparent for contributions due to $k=2$ and $k=32$. Hence, while the response function is dominated by frequency $\omega_{1}$, in fact many modes contribute to this behavior. 

However, because the normal-mode spectrum exhibits nonlinear dispersion, individual contributions tend to exhibit resonant frequencies lower than $\omega_{1}$. For example, while differences are small for long-wavelength modes, over longer timescales interference can become destructive. In Fig. \ref{K*1}, comparison between $k=2$ and $k=32$ contributions demonstrate a slight phase shift which accumulates with increasing correlation time $\tau$. The observed phase shifts track very closely the expected shifts based on the normal mode spectrum $\omega_{k}=|2 \sin\left({\pi k \over N}\right)|$. Hence, for $k=128$ and $k=192$, the normal-mode dispersion is strongly nonlinear, and $\omega_{k}-\omega_{k-q}$ is substantially smaller than $\omega_{1}$. The beats for these contributions are quickly not in phase with the dominant behavior at frequency near $\omega_{1}$. This phenomenon in part explains the contributions to $\tilde{K}^{T}(\omega)$ for frequencies in the range $|\omega| < \omega_{1}$ shown in Figs. \ref{RFT}-\ref{IFT}.

This analysis makes it clear why the dominant resonant behavior corresponds to frequency $\omega_{1}$. It also explains the primary reason for the decrease in $\tilde{K}_{q}(\tau)$ for relatively short timescales. Specifically, individual contributions have beat frequencies $\omega_{k}-\omega_{k-q}$ that differ from $\omega_{1}$ and hence eventually lead to a loss of phase coherence. This is not an effect related to scattering of the normal modes, but rather purely a harmonic wave-interference phenomenon. Hence, at least for the values of $\epsilon$ used in the simulations reported here, the dominant behavior can be understood as being related to nonlinear dispersion rather than to anharmonic scattering effects.

However, scattering and anharmonicity do play a notable and important role. This has been already demonstrated by comparison of the response $\tilde{K}_{q}(\tau)$ for different values of $\epsilon$. Resonant behavior tends to persist for longer periods of time for smaller values of $\epsilon$. This suggests scattering effects contribute to loss of phase coherence, with the effect becoming more pronounced as $\epsilon$ increases, much as would be expected. 

\begin{figure}
\begin{centering}
\includegraphics[width=0.5\textwidth]{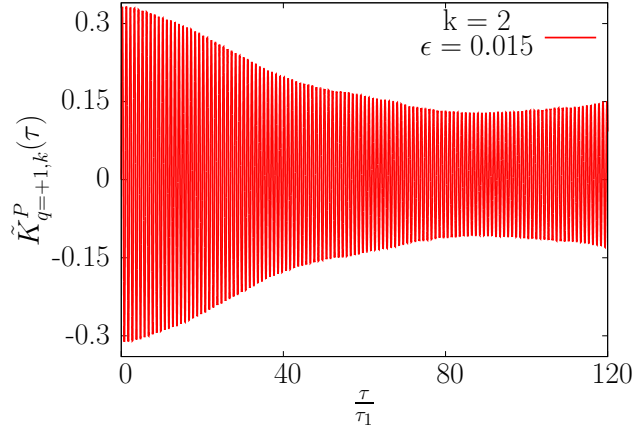} 
\caption{
Partial response function $\tilde{K}^{P}_{q=+1,k}(\tau)$ with $k=2$ obtained for $\epsilon=0.015$ and a system with $N=512$ particles.The time $\tau$ is scaled by the longest vibrational period in the system $\tau_{1}={\pi \over \sin{\pi \over N}}$. 
}
\label{K*2_0.015}
\end{centering}
\end{figure}

\begin{figure}
\begin{centering}
\includegraphics[width=0.5\textwidth]{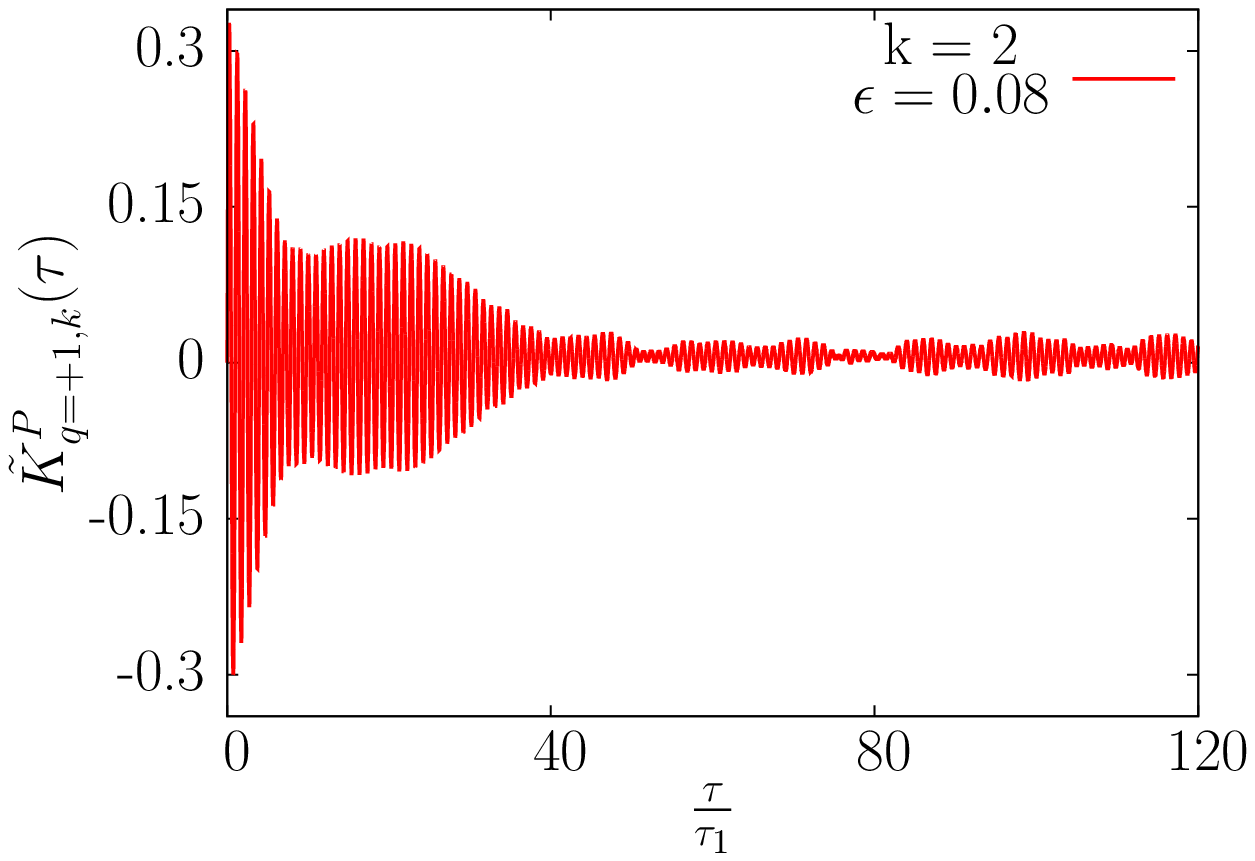} 
\caption{
Partial response function $\tilde{K}^{P}_{q=+1,k}(\tau)$ with $k=2$ obtained for $\epsilon=0.08$ and a system with $N=512$ particles.The time $\tau$ is scaled by the longest vibrational period in the system $\tau_{1}={\pi \over \sin{\pi \over N}}$. 
}
\label{K*2_0.08}
\end{centering}
\end{figure}

Evaluation of how the partial response function $\tilde{K}^{P}_{q=+1,k}(\tau)$ depends on $\epsilon$ and $k$ over longer simulation times demonstrates further that anharmonicity plays a role. For long wavelength modes, coherence is maintained for many oscillation periods, with $\epsilon$ acting to decrease the timescale for resonant behavior. This can be seen by comparing $\tilde{K}^{P}_{q=+1,k}(\tau)$ results in Fig \ref{K*2_0.015} and Fig \ref{K*2_0.08} for $k=2$. The results in Fig. \ref{K*2_0.015} were computed using $\epsilon=0.015$ and hence very low levels of anharmonicity. Even after $120$ periods $\tau_{1}$, coherence is still quite strong. By contrast, with $\epsilon=0.08$, the same quantity shows a substantial decay at around $40$ periods $\tau_{1}$.

\begin{figure}
\begin{centering}
\includegraphics[width=0.5\textwidth]{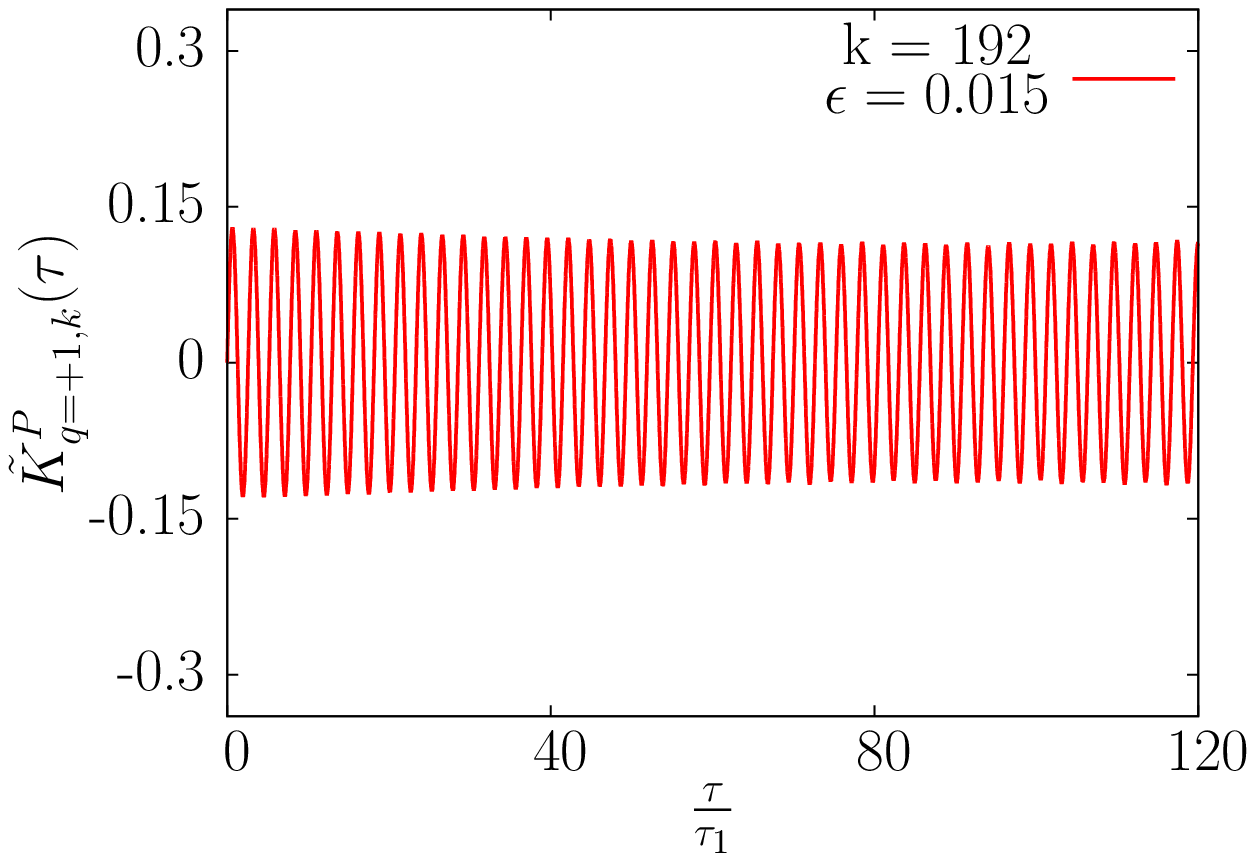} 
\caption{
Partial response function $\tilde{K}^{P}_{q=+1,k}(\tau)$ with $k=192$ obtained for $\epsilon=0.015$ and a system with $N=512$ particles.The time $\tau$ is scaled by the longest vibrational period in the system $\tau_{1}={\pi \over \sin{\pi \over N}}$. 
}
\label{K*192_0.015}
\end{centering}
\end{figure}

\begin{figure}
\begin{centering}
\includegraphics[width=0.5\textwidth]{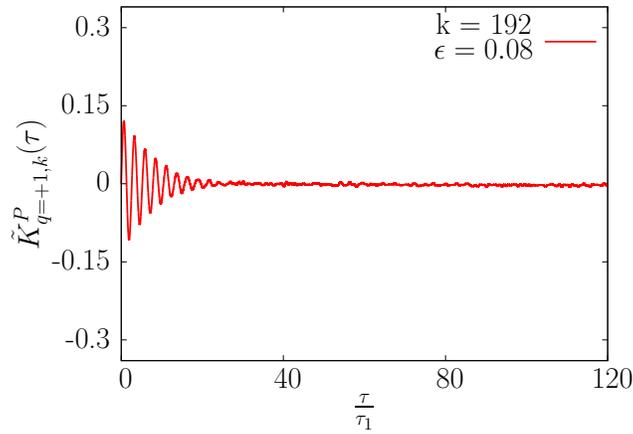} 
\caption{
Partial response function $\tilde{K}^{P}_{q=+1,k}(\tau)$ with $k=192$ obtained for $\epsilon=0.08$ and a system with $N=512$ particles.The time $\tau$ is scaled by the longest vibrational period in the system $\tau_{1}={\pi \over \sin{\pi \over N}}$. 
}
\label{K*192_0.08}
\end{centering}
\end{figure}

For larger values of $k$, the characteristic frequencies for the resonant behavior are sharply decreased as explained earlier. However, resonance is still clearly important. 
In Fig. \ref{K*192_0.015} for $k=192$ and $\epsilon=0.015$, resonant behavior is obtained beyond $120$ periods $\tau_{1}$. The oscillations themselves have a period substantially longer than $\tau_{1}$. The observed period is in good agreement with what is expected from the normal-mode dispersion. For $\epsilon=0.08$, scattering is evidenced by quite rapid loss of phase coherence. This is demonstrated in Fig. \ref{K*192_0.08}. 


\begin{figure}
\begin{centering}
\includegraphics[width=0.5\textwidth]{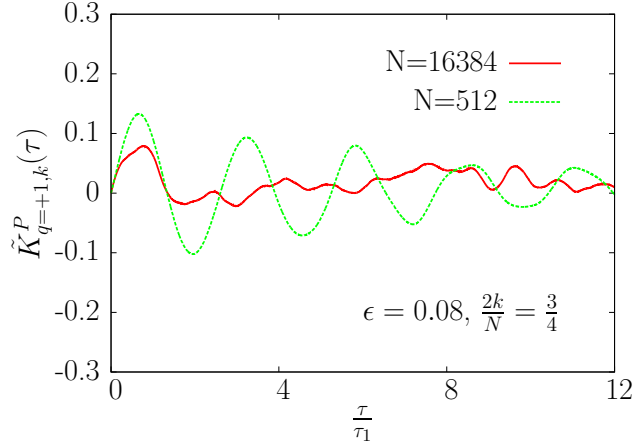} 
\caption{
Dependence of the partial response function $\tilde{K}^{P}_{q=+1,k}(\tau)$  for $\epsilon=0.08$ on chain length $N$. Comparison is made between $N=512$ and $N=16384$. The value of $k$ is given by ${2k\over N} = {3 \over 4}$. For $N=512$, this corresponds to $k=192$, and hence corresponds to the results in Fig. \ref{K*192_0.08}. For the longer chain with $N=16,384$,  $k=6144$. The period $\tau_{1}$ for scaling times on the horizontal axis is determined by the normal mode frequencies for the corresponding chain length $N$. For the $N=16,384$ results, numerical ensemble averaging used $2,000$ independent initial conditions with $3 \times 10^{6}$ MD steps each simulation.
}
\label{K*192_2}
\end{centering}
\end{figure}

While resonance is exhibited for all values of $\epsilon$ and $N$ simulated, analysis of the partial response functions $\tilde{K}^{P}_{q=+1,k}(\tau)$ show strong dependence on chain length $N$. As chain length increases, the beat frequencies $\omega_{k}-\omega_{k-1}$ decrease and the corresponding beat period $\tau={2 \pi \over \omega_{k}-\omega_{k-1}}$ increases, becoming significantly larger than $\tau_{1}$ as the zone edge is approached. Therefore,  phase coherence must be maintained for a much longer time for longer chains if resonant behavior is to be observed. We find that for small $k$, resonant behavior is present for $N=16384$, leading to the very comparable response functions for $N=512$ and $N=16,384$. However, analysis of the partial response $\tilde{K}^{P}_{q=+1,k}(\tau)$ for larger $k$ values shows a breakdown in resonant behavior. Therefore, while resonance is always observed in the response functions, the beginning of diffusive behavior can be seen for large values of $N$ and $k$. In Fig. \ref{K*192_2}, we show the comparison between $\tilde{K}^{P}_{q=+1,k}(\tau)$ for $N=512$ and $N=16,384$ chains. In both cases ${2k\over N} = {3 \over 4}$ and $\epsilon=0.08$ were chosen. For the $N=512$ chain, this corresponds to $k=192$ which was already shown in Fig. \ref{K*192_0.08}. What is evident from Fig. \ref{K*192_2} is that resonant behavior is quite pronounced for $N=512$, but for $N=16384$ scattering has destroyed phase coherence. What is important to remember here is that the beat period is longer for the $N=16,384$ chain by a factor of $32$ in comparison to the $N=512$ chain. Therefore, phase coherence between normal modes with frequencies $\omega_{k}$ and $\omega_{k \pm 1}$ must be maintained for a much longer time for the $N=16384$ chain in order to observe resonant behavior.

The picture above is consistent with the idea of ballistic transport at short time and length scales, and diffusive transport at longer time and length scales. It is clear that increasing chain length $N$ will eventually result in partial response functions $\tilde{K}^{P}_{q=+1,k}(\tau)$ exhibiting diffusive behavior for increasingly small values of $k$. Eventually, the complete response function $\tilde{K}_{q=+1}(\tau)$ will show behavior more consistent with diffusive transport. However, for very small $k$ values, we expect that as long as scattering is not strong enough to scatter at a rate comparable to ${\omega_{k} \over 2 \pi}$, at least some modes will always contribute  in a manner consistent with ballistic transport. This perspective is consistent with the observation of anomalous thermal transport in FPUT chains. 

Resonant behavior in the response functions depends on long lifetimes for the normal modes. Specifically, when a significant number of ensemble members exhibit incoherence, the response functions tend to decay to zero. If resonance is to be observed, the lifetime of normal modes must be greater than the period associated with the primary beat frequencies, which in the cases examined here this time is ${2 \pi \over \omega_{1}}$ . Hence, understanding scattering and loss of phase coherence is critical for understanding the response functions.
It is well-known that resonant three-wave scattering, which tends to dominate in three-dimensional solids, does not exist in one-dimensional FPUT chains\cite{Onorato:2015aa}. However, resonant four-wave scattering does exist\cite{BUSTAMANTE2019437}, and in smaller chains thermalization routes have been demonstrated for six-wave resonant scattering\cite{Onorato:2015aa}. Four wave scattering has been proposed to lead to nonlinear frequency shifts and also Umklapp-type scattering\cite{Onorato:2015aa}. To directly explore the scattering of the harmonic normal modes, we have computed the time-correlation function $C_{k}(\tau)=\langle P_{k}(\tau) P_{-k}(0) \rangle$.  This quantity can be Fourier transformed, 
\begin{equation}
\tilde{C}_{k}(\omega) = \int_{0}^{T_{max}} C_{k}(\tau) e^{-i \omega \tau} d\tau
\end{equation}
to explore the spectral properties. For harmonic behavior, we expect sharp peaks at frequencies $\omega=\omega_{k}$. However, scattering effects will lead to frequency shifts and peak broadening. 

\begin{figure}
\begin{centering}
\includegraphics[width=0.5\textwidth]{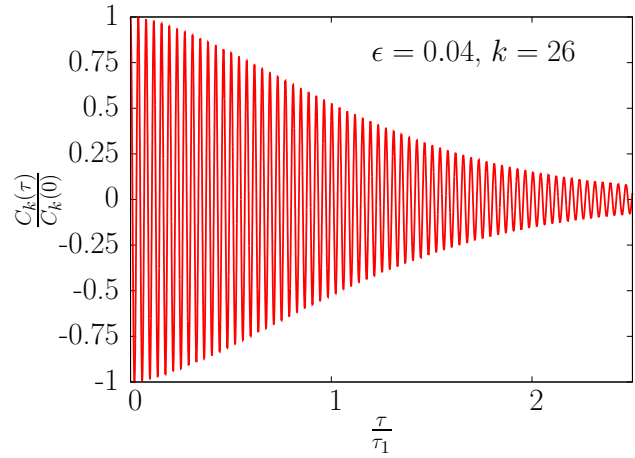} 
\caption{The quantity $C_{k}(\tau)$ normalized by the value at $\tau=0$ plotted as a function of time for $k=26$, $\epsilon=0.04$, and $N=512$. The ensemble used included $160$ independent initial conditions each integrating $1.5 \times 10^{6}$ MD steps. These ensemble details apply also to the data in Figs. 17-18.
}
\label{C26_eps0.04}
\end{centering}
\end{figure}

\begin{figure}
\begin{centering}
\includegraphics[width=0.5\textwidth]{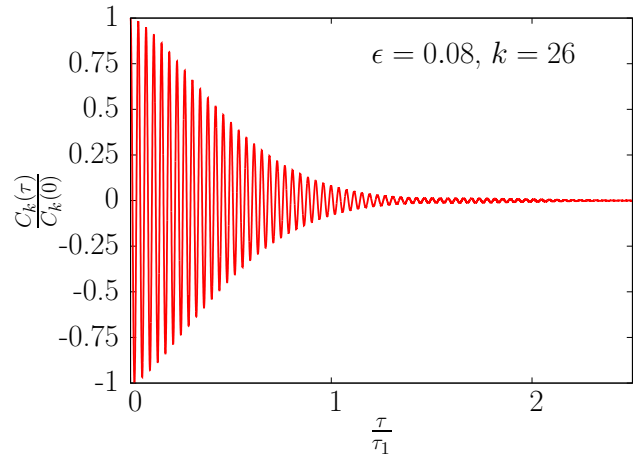} 
\caption{The quantity $C_{k}(\tau)$ normalized by the value at $\tau=0$ plotted as a function of time for $k=26$, $\epsilon=0.08$, and $N=512$.
}
\label{C26_eps0.08}
\end{centering}
\end{figure}

We have computed $C_{k}(\tau)$ and its Fourier transform $\tilde{C}_{k}(\omega)$ for modes with $k=26$ and $k=192$ for the $N=512$ chain. 
The results we have obtained qualitatively show that mode coherence persists across the spectrum for any value of $\epsilon$. This demonstrates that the normal modes themselves are coherent through many periods. As expected, coherence lifetime depends strongly on the value of $\epsilon$. In Fig. \ref{C26_eps0.04} we show $C_{k}(\tau) \over C_{k}(0)$ for $k=26$ and $\epsilon=0.04$. Comparison with the same quantity computed in conditions with $\epsilon=0.08$ shown in Fig. \ref{C26_eps0.08} demonstrates more rapid loss of coherence across the ensemble. However, what is surprising is that the rate of the loss of coherence appears to approximately depend linearly on $\epsilon$. If we recall that normal-mode amplitudes should vary with $\epsilon$ as $Q_{k} \sim \epsilon$ (see Eq. \ref{init} with the assumption that $\alpha$ is fixed), one would expect cubic anharmonic scattering rates to scale as $\epsilon^{2}$. This result can be demonstrated from Eq. 7 in Ref. \cite{Onorato:2015aa}. Similarly, as shown in Ref. \cite{Onorato:2015aa}, six-wave scattering, found to be dominant for short chains, leads to thermalization rates which scale as $\epsilon^{8}$. Moreover, four-wave scattering should lead to thermalization rates which scale as $\epsilon^{4}$\cite{Onorato:2015aa}. We have not yet found any explanation for the approximately linear dependence on $\epsilon$ demonstrated in our calculations.

\begin{figure}
\begin{centering}
\includegraphics[width=0.5\textwidth]{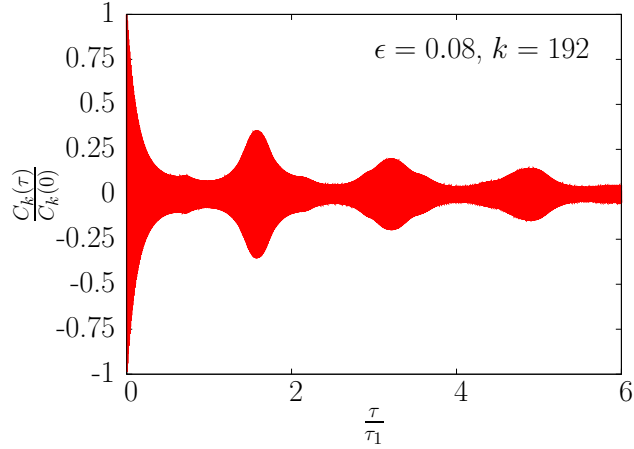} 
\caption{The quantity $C_{k}(\tau)$ normalized by the value at $\tau=0$ plotted as a function of time for $k=192$, $\epsilon=0.08$, and $N=512$.
}
\label{C192_eps0.08}
\end{centering}
\end{figure}

\begin{figure}
\begin{centering}
\includegraphics[width=0.5\textwidth]{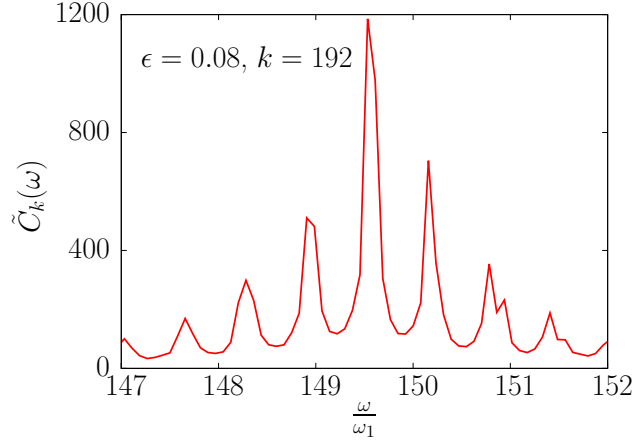} 
\caption{The quantity $\tilde{C}_{k}(\omega)$  plotted as a function of $\omega$ for $k=192$, $\epsilon=0.08$, and $N=512$.
}
\label{ft192_eps0.08}
\end{centering}
\end{figure}

For large values of $k$, we find the interesting result that phase-coherence itself can be periodic. In Fig. \ref{C192_eps0.08}, we show  $C_{k}(\tau)$ for $\epsilon=0.08$ and $k=192$. As can be seen, the oscillation period $\tau_{192}$ for this short-wavelength mode is much shorter than $\tau_{1}$. Hence, on the scale shown, the periodicity cannot be easily discerned. However, over longer times, phase coherence itself exhibits periodic behavior over much longer timescales. The Fourier transform $\tilde{C}_{k}(\omega)$ of this result is shown in Fig. \ref{ft192_eps0.08}. First we note that $\omega_{192}  \approx 150.6 \omega_{1}$, whereas the largest peak value in the Fourier transform occurs at $\omega \approx 149.5 \omega_{1}$. This suggest some nonlinear frequency shift possibly of the kind proposed to result from resonant four-wave scattering\cite{Onorato:2015aa}. More interesting is the presence of
multiple peaks spaced almost exactly by the difference $0.625 \omega_{1}$. We have not been able to associate this frequency with any features of the normal-mode spectra for $N=512$ chains.  One possibility is that this is the result of
the presence of so-called q-breathers demonstrated to exist in FPUT chains\cite{Flach:2005,FLACH_2007,Flach:2006wq}.  The basic idea is that energy in q-breathers is localized in reciprocal space, and can exhibit periodic recurrence phenomenon. It is possible that the frequency spacing between different peaks in Fig. \ref{ft192_eps0.08} might correspond to the frequency associated with the recurrence of q-breathers. We have seen this behavior in other normal modes as well, but the effect is not as dramatic for longer wavelength modes. However, the observed recurrence behavior may explain the fact that response functions $\tilde{K}_{q=+1} (\tau)$ show an initial decrease followed by an increase at later times. However, another contributing factor is likely that phase shifts due to nonlinear dispersion accumulate over time, so that after an initial decrease in $\tilde{K}_{q=+1} (\tau)$, coherent behavior might be realized again at a later time.  

\section{Conclusions}

The results here demonstrate the validity and usefulness of applying response functions to heat transport problems. 
Rather than starting from an assumption of diffusive or ballistic transport, a complete description of transport across all regimes can be obtained. In the $\alpha$-FPUT model, ballistic resonance is found to be the dominant transport mechanism. Furthermore, an analysis based on the normal-mode coordinates has demonstrated that beat frequencies $\omega_{k}-\omega_{k-q}$ and reinforcement across the spectrum is responsible for ballistic resonance and the resulting wave-like transport. While diffusive transport was not observed in the results presented here, detailed analysis across the normal-mode spectrum demonstrates the onset of diffusive behavior especially at higher frequencies and shorter wavelengths.

Previous reports of ballistic resonance in the $\alpha$-FPUT model have centered on the concept of mechanical vibrations which are excited due to local thermal expansion \cite{Kuzkin:2020tr}.  Subsequent work demonstrated that ballistic resonance does not occur in the  $\beta$-FPUT model due to the lack of thermal expansion with only quartic anharmonicity\cite{Korznikova:2020te}. If fact, in $\beta$-FPUT model, local thermal expansion does not occur. Nevertheless, resonant, wavelike behavior is still observed in the $\beta$-FPUT model. In fact, resonant behavior in the $\beta$-FPUT model was observed even in the case of a purely harmonic system with $\beta=0$. In both of these articles, continuum-level differential equations were used to describe the results \cite{Kuzkin:2020tr,Korznikova:2020te}.

It is important to note that the calculations reported previously\cite{Kuzkin:2020tr}, which highlighted the role of mechanical vibrations, used rather different initial conditions in contrast to the direct excitations simulated here based on Eq. \ref{excite}. In particular, while our excitations scaled velocities to generate an excitation, the referenced work \cite{Kuzkin:2020tr} initialized a state with each particle starting at its equilibrium position but with an initial velocity. Consequently, in contrast to the excitations used here, the previous approach started each excited normal mode exactly in phase. Based on our finding of reinforcing beat frequencies, this suggests that indeed a ``mechanical'' vibration might indeed be generated with growing amplitude using their approach\cite{Kuzkin:2020tr}. The results of this article demonstrate that resonant behavior occurs without the mechanism of mechanical vibrations simply due to harmonic effects based on reinforcing beat frequencies. In particular, the fact that the response functions, computed in equilibrium, demonstrate this behavior shows that the coupled thermo-viscoelastic response cited previously\cite{Kuzkin:2020tr} is not required to explain ballistic resonance.

The results reported here highlight the fact that spatially non-uniform heat inputs tend to generate phase coherence between normal modes. This is evident from Eq. \ref{uharm}, which when applied to the Fourier components of an external source $\tilde{u}^{(ext)}_{q}$ show there must be phase coherence between modes $\pm k$ and $\mp k \pm q$ across the entire spectrum of harmonic normal modes. The same phase coherence is reflected in the Fourier components of the heat current $\tilde{J}_{q}$ after action of the external source. This is shown in Eq. \ref{currpq} and Eq. \ref{currmq}. In fact, we note that the expressions in Eqs. \ref{currpq}-\ref{currmq} are essentially identical to the local harmonic current first derived by Hardy\cite{Hardy:1963td}. Specifically, the local form of the current operator $\vec{J}(\vec{x})$ at position vector $\vec{x}$ is given by Hardy in Eq. 5.10-5.11 as\cite{Hardy:1963td},
\begin{equation}
\vec{J}(\vec{x}) = \sum_{\vec{k}} N_{\vec{k}}(\vec{x}) \hbar \omega_{\vec{k}} \vec{v}_{\vec{k}}
\end{equation}
where for clarity and simpicity, while this is written assume a three-dimensional system, the band index $s$ has been omitted, and the notation has been altered somewhat to be consistent with our work. The local distribution operator $N_{\vec{k}}(\vec{x})$ is then given by,
\begin{equation}
N_{\vec{k}}(\vec{x}) = {1 \over 2V}
\sum_{\vec{k}^{\prime}} \left[a_{\vec{k}}^{\dagger} a_{\vec{k}^{\prime}} e^{i(\vec{k}^{\prime}-\vec{k})\cdot \vec{x}} +
a_{\vec{k}^{\prime}}^{\dagger} a_{\vec{k}} e^{-i(\vec{k}^{\prime}-\vec{k})\cdot \vec{x}} 
\right ] \exp{\left[-{1 \over 4} | \vec{k} -\vec{k}^{\prime}|^{2} l^{2} \right]}
\end{equation}
in which $V$ is the system volume, $l$ is a length scale associated with a Gaussian localization function, and $a_{\vec{k}}^{\dagger}$, $a_{\vec{k}^{\prime}}$ are phonon creation and annihilation operators. The interpretation of $N_{\vec{k}}(\vec{x})$ as the density of phonons with wave vector $\vec{k}$ at point $\vec{x}$ is not entirely straightforward. Rather, if one takes $\vec{k}^{\prime}-\vec{k} = \pm \vec{q}$, then the relation between Hardy's local current and the Fourier components $\tilde{J}_{\pm q}$ defined by Eqs. \ref{currpq}-\ref{currmq} becomes clear. Then given the time dependence of operators, it is clear that Hardy's expression also involves slow oscillations with frequency $\omega_{\vec{k}}-\omega_{\vec{k}^{\prime}}$, a point which was noted in his paper\cite{Hardy:1963td}. Therefore, it is clear that in a local representation of current, the phase relation between different modes determines the behavior over short timescales until scattering processes act to destroy the coherence. This picture of transport is somewhat different from the standard picture that requires a nonequilibrium distribution of phonons. Nevertheless, we believe that the picture presented here is consistent with derivations in Ref. \cite{Hardy:1963td} for the local current operator. This also shows that nonequilibrium states, characterized by nonuniform energy distributions, can arise due to phase coherence between large numbers of phonon modes, and not only due to phonon occupations that differ from the equilibrium Bose-Einstein distribution. 

The decay of resonant behavior has two origins. First, due to nonlinear dispersion, coherence between beats decreases with time. This is a purely harmonic effect which explains previous results in the  $\beta$-FPUT model when the anharmonic interaction parameter $\beta=0$ and evolution is entirely harmonic\cite{Korznikova:2020te}, but also other cases when anharmonicity is included in the FPUT model\cite{Kuzkin:2020tr,Korznikova:2020te}.  While this mechanism does cause decay in the resonant behavior, it is unrelated to anharmonic interactions usually studied as a mechanism to attain classical equipartition\cite{Onorato:2015aa}. However, this effect does lead the system towards a uniform energy distribution in real space, and the eventual elimination of local temperature gradients. The second mechanism for the decay of resonant behavior is due to anharmonicity. In the results here with $\alpha>0$, normal mode scattering leads to a loss of phase coherence in the ensemble. This mechanism is also responsible for the thermalization of a non-equilibrium ensemble.

It is interesting to speculate on how these observations might be relevant to two- and three-dimensional materials. In the original paper outlining the methodology of thermal response functions\cite{Fernando_2020}, we demonstrated that ballistic transport was clearly apparent at short length scales and low temperatures in a three-dimensional lattice with Lennard-Jones interactions. However, no indication of oscillatory transport was found. Hence, some signatures of ballistic transport, perhaps more specifically ballistic resonance, may be harder to observe in three-dimensional systems. We believe that the fact that three-dimensional materials have a much more complex normal-mode spectrum, and hence less of a tendency for reinforcing beat frequencies, is likely responsible for these differences. Specifically, in one-dimensional systems like the FPUT model, for a very significant part of the spectrum $\omega_{k}-\omega_{k-q} \approx \omega_{q}$ can be assumed, and hence beats tend to reinforce. By contrast, in a three-dimensional lattice, beat phenomena likely exist, but due to the complex spectrum, beats should act more incoherently. The same considerations will apply to two-dimensional materials. As a result, ballistic resonance effects seen in one-dimensional chains might not be present or at least as striking in higher-dimensional systems.


We plan to apply the approach developed here to more realistic materials to test some of these ideas. As mentioned previously, there have been recent reports of second sound in graphite\cite{Ding_2022} and even bulk germanium\cite{Beardo_2021} which might present interesting test cases. Finally, we suggest that computation of response functions might be a complementary approach to other methods including solution of the Peierls-Boltzmann equation (PBE). In fact, in a few recent publications analogous approaches have been developed using solutions of the PBE\cite{Hua:2020wt,Allen:2018wa}. For example, the concept of ``thermal susceptibility'' developed by Allen and Perebeinos\cite{Allen:2018wa} is identical to what we have called thermal response. It should be possible to take a first-principles approach to the computation of scattering rates as is typically implemented in PBE studies of bulk conductivity\cite{Lindsay_2016}, but instead to determine timescales for the loss of phase coherence. It should be straightforward to extend the approach to the low-temperature regime where Bose-Einstein statistics should apply, and thereby develop an approach to understand thermal response at very low temperatures where ballistic transport, second sound, and phonon hydrodynamics are particularly relevant. We note that if these functions could be determined over a wide range of conditions, transport across ballistic, diffusive, and intermediate regimes could be computed. Moreover, while the response functions clearly involve a substantial amount of data, they actually represent the integrated contributions of the entire spectrum, and thereby might represent an efficient, compact, and accurate approach towards modeling thermal transport.

\section{Acknowledgements}
We would like to thank Prof. Philip Allen for valuable discussions, as well as providing several suggestions and corrections for improvements to the text. We also benefited greatly from reading his closely-related works, both for additional presentation of the concept of thermal susceptibility and transport in one-dimensional chains. We would also like to thank the referees for pointing out several errors in the text.

\newpage

\bibliographystyle{unsrt}

\end{document}